\documentclass[10pt, journal, letterpaper, twocolumn]{IEEEtran}
\usepackage[dvips]{epsfig}

\usepackage{amsfonts}
\usepackage{amsmath}
\usepackage{amssymb}
\usepackage{pslatex}
\usepackage{dblfloatfix}

\begin{document}
%
\title{Bright single photon emission from a quantum dot in a circular Bragg grating microcavity}
\author{Serkan~Ate\c{s}, Luca~Sapienza, Marcelo~Davan\c co, Antonio~Badolato, and Kartik~Srinivasan
\thanks{S. Ate\c{s}, L. Sapienza, M. Davan\c co, and K.
Srinivasan are with the Center for Nanoscale Science and Technology,
National Institute of Standards and Technology, Gaithersburg, MD
20899-6203, USA (e-mail: serkan.ates@nist.gov; kartik.srinivasan@nist.gov)}\\
\thanks{A. Badolato is with the Department of Physics
and Astronomy, University of Rochester, Rochester, NY 14627}
\thanks{S. Ate\c{s} and M. Davan\c co are also with the Maryland
Nanocenter, University of Maryland, College Park, MD 20742}
\thanks{L. Sapienza's present address is School of Engineering and Physical
Sciences, Heriot-Watt University, Edinburgh, EH14 4AS, UK}}


\maketitle

\begin{abstract}
Bright single photon emission from single quantum dots in suspended
circular Bragg grating microcavities is demonstrated. This geometry
has been designed to achieve efficient ($>50~\%$) single photon
extraction into a near-Gaussian shaped far-field pattern, modest
($\approx$10x) Purcell enhancement of the radiative rate, and a
spectral bandwidth of a few nanometers.  Measurements of fabricated
devices show progress towards these goals, with collection
efficiencies as high as $\approx10~\%$ demonstrated with moderate
spectral bandwidth and rate enhancement. Photon correlation
measurements are performed under above-bandgap excitation (pump
wavelength = $780$ nm to $820$ nm) and confirm the single photon
character of the collected emission. While the measured sources are
all antibunched and dominantly composed of single photons, the
multi-photon probability varies significantly. Devices exhibiting
tradeoffs between collection efficiency, Purcell enhancement, and
multi-photon probability are explored and the results are
interpreted with the help of finite-difference time-domain
simulations. Below-bandgap excitation resonant with higher states of
the quantum dot and/or cavity (pump wavelength = $860$\,nm to
$900$\,nm) shows a near-complete suppression of multi-photon events
and may circumvent some of the aforementioned tradeoffs.

\end{abstract}

\section{Introduction}
\label{sec:Intro}

Single epitaxially-grown In$_x$Ga$_{1-x}$As quantum dots (QDs) have
generated significant interest as potentially bright and stable
single photon sources for quantum information processing
applications~\cite{ref:Shields_NPhot,ref:Michler,ref:Santori}.  In
principle, the maximum achievable single photon rate is limited only
by the spontaneous emission lifetime ($\tau_{\text{sp}}$) of the QD,
so that for a single InGaAs QD with a typical
$\tau_{\text{sp}}\approx$1 ns in bulk, a single photon rate
$R\approx1/\tau_{sp}$ approaching 1 GHz may be possible. Still
faster rates can be achieved through Purcell enhancement of the QD
radiative rate~\cite{ref:Gerard1}. In practice, however, the
available single photon rates are not necessarily limited by the
generation rate, but by the efficiency with which the single photons
are funneled into a useful collection channel (e.g., low divergence
angle far-field emission). This well-known
problem~\cite{ref:Benisty3,ref:Barnes2,ref:Gerard3} is fundamentally
due to the high-refractive index contrast between the GaAs material
in which the QD is embedded (refractive index $n\approx3.4$) and the
surrounding air ($n$=1). This strongly limits the amount of
free-space emission exiting the semiconductor because of total
internal reflection. Even with high numerical aperture (NA) optics,
a maximum theoretical collection efficiency $<1~\%$ is expected.

This issue has been addressed by a number of groups.  Solid
immersion
lenses~\cite{ref:Zwiller_Bjork_JAP,vamivakas.nano.letters.7.2892}
provide a broadband and versatile approach to directly increase the
collection efficiency. Alternatively, nanofabricated photonic
structures such as optical
microcavities~\cite{ref:Strauf_NPhot,ref:Solomon,ref:Toishi} and
waveguides\cite{ref:Claudon,ref:Davanco_WG} provide confined or
guided modes into which the QD can radiate.  Ideally, the QD will
dominantly radiate into a single mode of the structure (high
spontaneous emission coupling factor $\beta$), which can then be
efficiently out-coupled in the far-field depending on its emission
pattern and divergence angle.  Recent
work~\cite{ref:Strauf_NPhot,ref:Claudon} has demonstrated
efficiencies in excess of 50$~\%$ into the first optical element of
a fluorescence microscopy setup.  The microcavity and waveguide
approaches each offer their own advantages - microcavities
supporting high quality factor ($Q$) optical modes provide radiative
rate enhancement, albeit over a narrow spectral band (the width of
the cavity mode), while waveguides are broadband devices but
typically do not offer rate enhancement (unless they operate in a
slow-light regime). Broadband operation avoids the need for precise
spectral alignment of the QD emission line to a photonic resonance
and allows for efficient spectroscopy of multiple spectrally
separated states of a single QD. On the other hand, radiative rate
enhancement increases the maximum rate at which single photons can
be generated by the QD, and can improve the indistinguishability of
the photons~\cite{ref:Varoutsis_PRB05,ref:Santori2}.

\begin{figure}[t]
\centerline{\includegraphics[width=\linewidth]{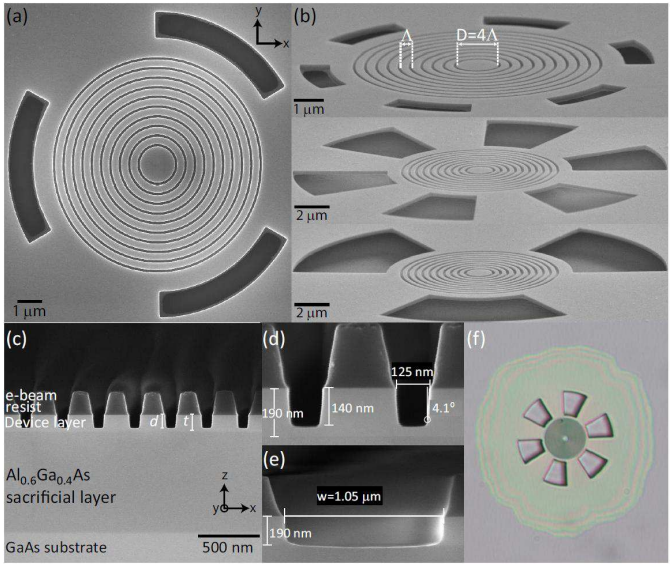}}
 \caption{(a) Top view and (b) angled view scanning electron microscope (SEM)
 images of the bullseye cavities with different undercut access geometries.
 (c)-(e) Cross-sectional SEM images of the bullseye cavities after the GaAs dry etch but before the HF undercut.
The grating trenches are etched to a depth $d$ that is below the
QD-containing layer (located at $t/2$) but smaller than the GaAs
thickness ($t$). In comparison, the access areas outside of the
grating are fully etched through the GaAs layer, allowing the
devices to be undercut. (f) Optical microscope image of a fully
processed device.} \label{fig:Figure1}
\end{figure}

In ref.~\cite{ref:Davanco_BE}, we proposed and demonstrated a
microcavity geometry for efficient extraction of photons from single
QDs.  This 'bullseye' geometry, shown in Fig.~\ref{fig:Figure1},
consists of a series of partially etched grooves in a suspended GaAs
membrane containing a layer of InAs QDs.  In comparison to many of
the bright single photon geometries demonstrated thus
far~\cite{ref:Solomon,ref:Strauf_NPhot,ref:Claudon}, which are based
on vertically-oriented micropillars or nanowires, the bullseye
geometry is planar, thereby requiring a relatively simple
fabrication process. The fundamental optical mode of this cavity has
a small volume $V_{\text{eff}}\approx1.3(\lambda/n)^3$, but a
relatively low $Q=200$.  This allows for both modest Purcell
enhancement (maximum value 12$\times$ predicted in theory) and
moderate spectral bandwidth (few nm for QDs emitting at
$\lambda\approx940$ nm). In addition, the geometry supports a
far-field emission pattern that is near-Gaussian and predominantly
directed vertically above the sample surface.  Collection
efficiencies higher than $50~\%$ (for NA $\gtrsim0.42$) are
predicted theoretically, and a measured efficiency of 10$~\%$ was
reported in ref.~\cite{ref:Davanco_BE}.

In this paper, we present further developments of the bullseye
cavity, with a focus on better understanding its potential in single
photon source applications. A key question left unanswered in
ref.~\cite{ref:Davanco_BE} is the degree to which multi-photon
probability is suppressed in the out-coupled emission from the
device, given the fact that the QD lines are situated on top of a
broad cavity mode which also effectively out-couples other (non
single photon) emission from multi-excitonic and hybridized
QD-wetting layer
states~\cite{ref:Winger2009,ref:Laucht_Finely_2010}. Through photon
correlation measurements of a series of devices under above-bandgap
excitation (pump wavelength = $780$ nm to $820$ nm), we observe
tradeoffs between collection efficiency, Purcell enhancement, and
multi-photon probability.  These tradeoffs, and the reduction in
collection efficiency and Purcell enhancement relative to the
maximum predicted values, are interpreted with the help of
finite-difference time-domain simulations which focus on the role of
variations in the grating geometry and QD position and orientation
in influencing device performance.  The results indicate that if
adequate control of the QD position~\cite{ref:Hennessy3} and grating
dimensions can be exercised, the device behavior may be tuned
according to the application at hand (for example, limiting Purcell
enhancement in exchange for single photon purity).  Ideally, such
tradeoffs would not be necessary, and as a first step towards
improving this, we pursue below-bandgap excitation resonant with
higher states of the QD and/or cavity (pump wavelength = $860$ nm to
$890$ nm). These measurements show a near-complete suppression of
multi-photon events and may circumvent some of the aforementioned
limitations.

\section{Device Design}
\label{sec:design}

Our nanophotonic structure (Fig.~\ref{fig:Figure1}) consists of a
circular dielectric grating with radial period $\Lambda$ surrounding
a central region of radius $R=$2$\Lambda$, produced on a suspended
GaAs membrane. The GaAs slab of thickness $t$ supports single TE and
TM polarized modes (electric or magnetic field parallel to the slab,
respectively) at wavelengths near 980~nm. The grating is composed of
partially etched circular trenches that have width $w$, depth $d$
($t/2<d<t$), and are radially spaced by a period $\Lambda$. Quantum
dots are grown at the center of the GaAs membrane ($z=0$), and are
located randomly in the $xy$ plane.

This 'bullseye' geometry favors extraction of emission from QDs in
the central circular region, which would otherwise be trapped inside
the GaAs membrane. Indeed, a QD located in the central region
radiates dominantly into slab-guided waves due to total internal
reflection at the semiconductor-air interface. The role of the
circular grating is to scatter such guided waves into free-space,
preferentially upwards, towards the collection optic. Effective
light extraction can be achieved via a second order Bragg grating
for the slab modes (with period equal to the guided wave
wavelength), which provides a first order diffraction perpendicular
to the slab plane~\cite{ref:Hardy,ref:Baets1}. In~\cite{ref:Baets1},
linear gratings were developed to provide efficient coupling between
free-space beams and planar waveguides modes. In our case, a
circular grating geometry was a more natural choice because QD
emission in the GaAs slab was expected to be cylindrically
symmetric. We point out that similar circular geometries have been
employed for enhanced light extraction from light emitting
diodes~\cite{ref:su_APL_033105}, and for demonstrating annular Bragg
lasers~\cite{ref:Green}.

The grating period $\Lambda$ was chosen to approximately satisfy the
second-order Bragg condition, $\Lambda=\lambda_{QD}/n_{TE}$, where
$n_{TE}$ is the GaAs slab TE mode effective index. Because this
formula is valid in the limit of a weak grating, which is not the
case here (as the grating's etch depth is a considerable fraction of
the slab thickness), we use it as a starting point in an
optimization procedure which determines the correct period.

\begin{figure}[t]
\centerline{\includegraphics[width=8.5cm]{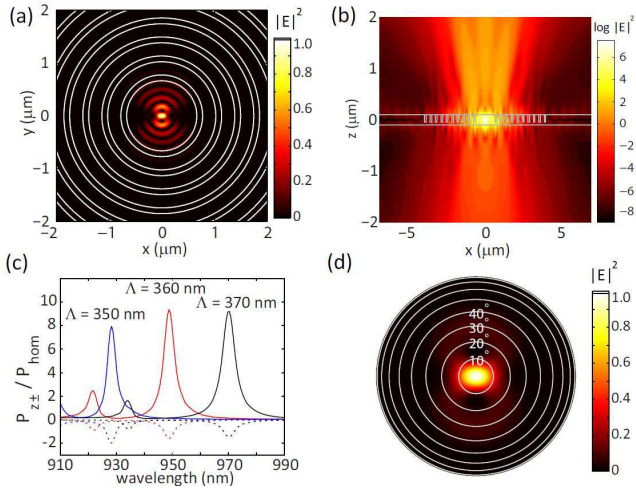}}
\caption{Electric field intensity in the (a) $xy$ and (b) $xz$
planes (log scale). (c) Calculated vertically extracted power as a
function of wavelength, normalized to the homogeneous medium
electric dipole power $P_{\text{hom}}$ for $d=0.70t$. Continuous
lines: upwards ($+z$) extraction; dotted: downwards ($-z$). (d)
Far-field polar plot for the cavity mode with $\Lambda=360$~nm.}
\label{fig:Figure2}
\end{figure}

Partial reflections at the grating towards the center of the
geometry lead to the formation of cavity resonances such as shown in
Fig.~\ref{fig:Figure2}(a), where the particular orientation of the
cavity mode is determined by the orientation of the radiating dipole
($x$ direction in this case). The large index contrast at the
trenches leads to strong reflections and out-of-slab-plane
scattering at the semiconductor-air interfaces, as apparent in
Fig.~\ref{fig:Figure2}(b). Resonances are centered at wavelengths
determined by the radius of the central region and the grating
geometry itself, which determines the overall phase of the reflected
waves. Figure~\ref{fig:Figure2}(c) shows the evolution of the
resonance center wavelength for three varying grating periods
$\Lambda$. In addition to the trench spacing, large differences in
propagation constants in the semiconductor and air regions of the
grating produce significant resonance spectral shifts with small
variations in trench width and depth.

The trench depth $d$ has a strong influence on the quality factor
($Q$) and vertical light extraction, as incomplete spatial overlap
between a trench and an incident slab-bound wave leads to both
coupling to radiating waves and lower modal reflectivity.
Preferential upwards vertical extraction results from the grating
asymmetry, as evidenced in Fig.~\ref{fig:Figure2}(c). This
trade-off between the quality factor $Q$, power extraction and power
extraction asymmetry, and far-field collimation due to variation in
the trench depth is discussed further in
Section~\ref{subsec:sim_discuss}. We note that our suspended,
asymmetric grating approach limits radiation into the substrate
without the need to oxidize the AlGaAs, bond the grating to a low
index layer\cite{ref:Green}, or utilize a deeply etched
geometry~\cite{ref:Strauf_NPhot,ref:Claudon}.

In Fig.~\ref{fig:Figure2}(c), the maximum upwards extracted powers
are $\approx10\times P_{\text{hom}}$ ($P_{\text{hom}}$ is the
emitted power for an electric dipole in a homogeneous GaAs medium),
an indication of Purcell radiation rate enhancement due to the
cavity~\cite{ref:Xu_99,ref:Vuckovic1}. Indeed, for the
$\Lambda=360$~nm structure the enhancement $F_p$ at the maximum
extraction wavelength ($\lambda_c = 948.9$~nm) is
$F_p=P_\text{tot}/P_\text{Hom}=11.0$, where $P_\text{tot}$ is the
total radiated power in all directions. This resonance has $Q=200$,
and its effective mode volume, calculated from the field
distribution, is $V_\text{eff}=1.3(\lambda_c/n)^3$ ($n$ is the GaAs
refractive index)~\cite{ref:Davanco_BE}. Thus, despite the
relatively low quality factor, the mode volume can be sufficiently
small to produce a significant Purcell factor.

Finally, the far-field of the emitted radiation is highly
directional, as evidenced in Fig.~\ref{fig:Figure2}(d). Here, the
emitted field for a dipole located at the bullseye center is nearly
gaussian, and mostly contained within a $20^{\circ}$ half-angle.
Indeed, in~\cite{ref:Davanco_BE}, we show that $\approx53~\%$ of the
total emission can be collected with a NA=0.42 optic, while
collection superior to 80$~\%$ of the total emission can be achieved
with NA$>0.7$. The dependence of the emitted field's spatial
distribution on geometric parameters is non-trivial, however, and is
discussed further in Section~\ref{subsec:sim_discuss}.

\subsection*{Design Procedure}

The design process used to generate Fig.~\ref{fig:Figure2} above
and the results to be discussed in Section \ref{subsec:sim_discuss}
consisted of a series of finite-difference time-domain simulations
that sought to maximize vertical light extraction near the QD's
expected s-shell emission ($\lambda_{QD}\approx940$~nm), by varying
$\Lambda$, $t$, and $w$. The structures were excited with a
horizontally oriented electric dipole at the bullseye center
$(x=0,y=0)$, representing an optimally placed QD. Total radiated
power, steady-state upwards and downwards emission, and
electromagnetic fields were then recorded at several wavelengths.
The dipole orientation was assumed to be aligned along the $xy$
plane, exciting only TE slab waves (self-assembled InAs QDs are
expected to have electric dipole on the GaAs slab plane).

Steady-state fields at a surface just above the GaAs slab were used
to calculate far-field patterns as in Fig.~\ref{fig:Figure2}(d).
The power $P_\text{coll}$ collected by an optic of numerical
aperture NA was calculated by integrating the far-field pattern over
the appropriate angular range, and multiplying by the power radiated
upwards.

\section{Experimental Results Under Above-Band Excitation}
\label{sec:expt_above_band}

\subsection{Experimental Setup}
\label{subsec:expt_setup}

\begin{figure*}[t]
\begin{center}
\begin{minipage}[c]{0.78\linewidth}
\includegraphics[width=0.95\linewidth]{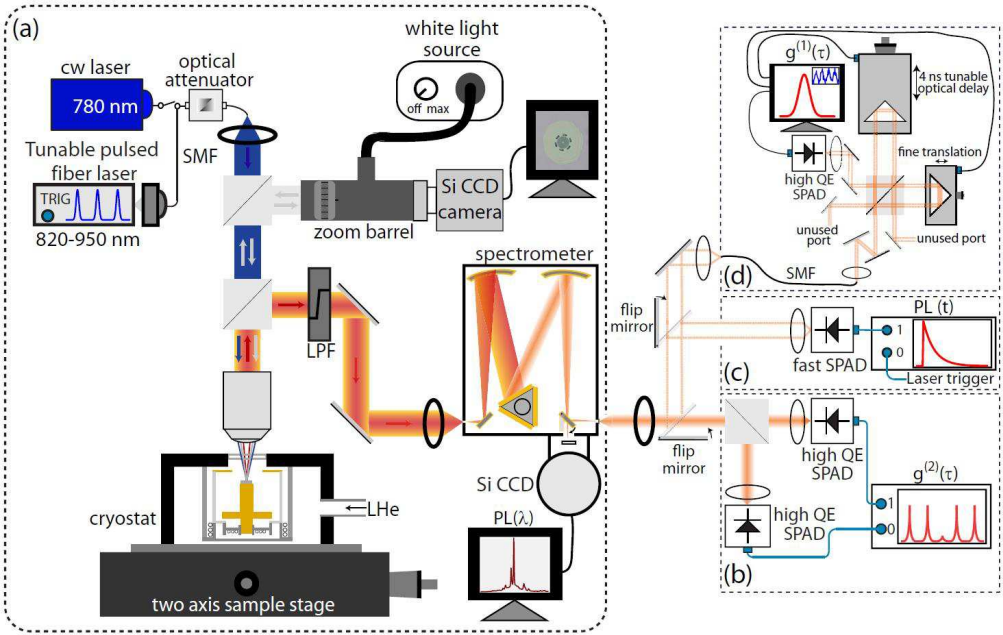}
\end{minipage}\hfill
\begin{minipage}[c]{0.19\linewidth}
\caption{Schematic of the experimental setup: (a) Confocal $\mu$-PL,
(b) Hanbury-Brown and Twiss photon correlation setup, (c)
time-resolved PL, and (d) Michelson interferometer. SMF: single mode
fiber, LPF: long pass filter} \label{fig:Figure3}
\end{minipage}
\end{center}
\end{figure*}

Figure~\ref{fig:Figure3} shows a detailed schematic view of several
experimental setups used within this work. The sample was mounted on
a cold finger and placed in a continuous flow liquid He cryostat
that sits on a two-axis nano-positioning stage. Two different
excitation sources were used to optically excite the QDs: (i) a CW
laser diode ($780$~nm) for above-band excitation, (ii) a tunable
pulsed fiber laser ($820$~nm to $950$~nm) for quasi-resonant
excitation. Spectral properties of the QD emission were investigated
via a low-temperature micro-photoluminescence ($\mu$-PL) setup
(Fig.\ref{fig:Figure3}(a)), where a single microscope objective
(NA\,=\,0.42) was used for both the illumination of the sample and
the collection of the emission. The collected signal is directed to
a 500\,mm focal length spectrometer either to record an emission
spectrum with a Si Charge-Coupled Device (CCD) or to filter a single
emission line for further investigations. For the second-order
correlation function $g^{(2)}(\tau)$ measurements, the spectrally
filtered emission (with a linewidth
$\approx70$~pm~$\approx$100~$\mu$eV) is directed to a Hanbury-Brown
and Twiss (HBT) type interferometer that consists of a 50/50
non-polarizing beamsplitter (NPBS) and two high quantum efficiency
(QE) single-photon counting avalanche diodes (SPADs), as shown in
Fig.~\ref{fig:Figure3}(b). The SPADs (peak QE\,=\,73~\% at 700\,nm and
QE\,=\,28~\% at 980\,nm) were connected to a time-correlated
single-photon counting module to create a histogram of photon
detection events. The dynamics of a single QD emission was measured
by using a time-correlated single-photon counting (TCSPC) technique,
which relies on measuring the time delay between an excitation pulse
and detection of an emitted photon by using a fast SPAD
(Fig.~\ref{fig:Figure3}(c), timing jitter = 50\,ps, QE\,=\,3~\% at
940\,nm). Finally, for first-order field correlation function
$g^{(1)}(\tau)$ measurements, a Michelson interferometer was
attached to the output of the spectrometer. As shown in
Fig.~\ref{fig:Figure3}(d), the setup consists of a 50/50 NPBS and two
retro-reflectors, one of which is mounted on a linear stage to
provide a coarse tunable optical delay up to 4\,ns and the other
retro-reflector is attached on a piezo actuator for a fine delay.
The interference fringes were recorded by a SPAD, which was attached
to one output port of the interferometer.

\subsection{Collection efficiency, time-resolved photoluminescence,
and photon antibunching}

\begin{figure}[t]
\centerline{\includegraphics[width=0.95\linewidth]{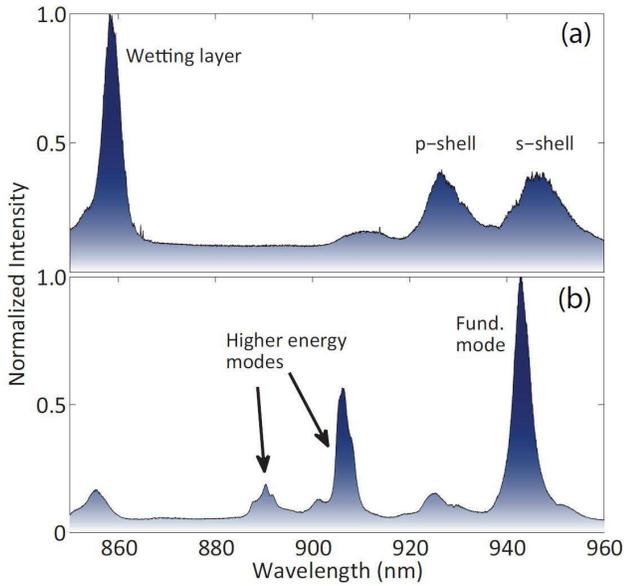}}
 \caption{(a) Low temperature broad range PL spectrum of QD-ensemble under high excitation power conditions.
S-shell, p-shell and wetting layer emission of the ensemble are
observed. (b) Low temperature broad range PL spectrum of a Bullseye
device which shows the fundamental mode and the higher energy modes
of the cavity. } \label{fig:Figure4}
\end{figure}

Initial characterization of the bare (unprocessed) QD sample has
been done under strong pump powers with an excitation energy above
the GaAs bandgap ($780$~nm), which excites all QDs within the
excitation spot. Figure~\ref{fig:Figure4}(a) shows a low
temperature wide range $\mu$-PL spectrum of the QD ensemble, where
clear s-shell and p-shell emission of the ensemble are observed
around $945$~nm and $927$~nm, respectively. The sharper peak around
$858$~nm is attributed to the wetting layer of the sample. In order
to limit the number of QDs coupled to a mode, the bullseye devices
are fabricated in a low QD density ($\approx$1$~\mu$m$^{-2}$)
portion of the wafer and are designed to have their fundamental
cavity mode spectrally aligned with the tail of the s-shell of the
QD ensemble. We focus on three devices, named as BE1, BE2, and BE3,
which have the same nominal grating period $\Lambda$ and central
diameter $D$, but differ due to slight variations in fabrication
across the chip (trench width and depth) and perhaps most
importantly, due to the varying QD location and orientation within
the devices (which is random). Figure~\ref{fig:Figure4}(b) shows
a similar wide range high power $\mu$-PL spectrum of device BE1,
which clearly shows the fundamental mode (FM) and the well-separated
higher energy modes of the cavity. The quality factor of the FM is
measured from the ratio of the emission energy to the linewidth as
$Q$\,=\,200.

\begin{figure}[t]
\centerline{\includegraphics[width=0.95\linewidth]{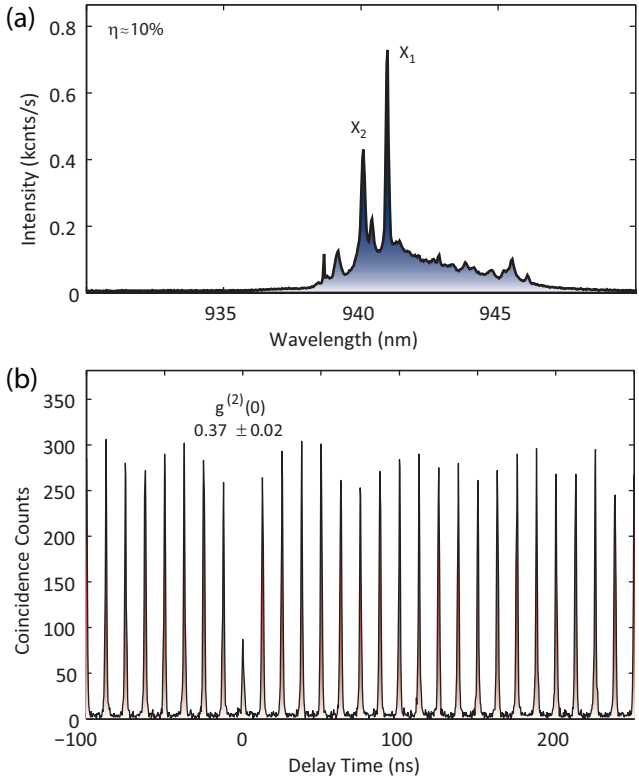}}
 \caption{(a) A typical low power $\mu$-PL spectrum of BE1 under pulsed GaAs excitation at a
temperature T\,=\,10\,K. Lines $X_{1}$ and $X_{2}$ at $940$~nm and
$941$~nm, respectively, are states of a single QD that yield a
collection efficiency $\eta$ of 10~$\%$. (b) The result of
second-order correlation function measurement on the $X_{1}$ line. A
clear suppression of the central peak with
$g^{(2)}(\tau)\,=\,0.37\,<\,0.5$ proves the single-photon nature of
the collected emission.} \label{fig:Figure5}
\end{figure}

Figure~\ref{fig:Figure5}(a) shows a low temperature (T\,=\,10\,K)
$\mu$-PL spectrum of BE1 taken under weak pulsed excitation of the
GaAs ($\lambda=820$~nm). This device is the same as that presented
in ref.~\cite{ref:Davanco_BE}, and displays bright emission lines
labeled X$_{1}$ and X$_{2}$ that sit on top of a broad background
due to the cavity mode emission, which is thought to be mainly fed
by the multi-exciton states of several nearby QDs
\cite{ref:Winger2009}. Both X$_{1}$ and X$_{2}$ emission lines have
almost 10~$\%$ collection efficiency $\eta$ at their saturation
powers, and time-resolved PL measurements performed on the X$_1$
line resulted in a lifetime as fast as 360$~\pm$6~ps, which is
likely indicative of Purcell rate enhancement, as discussed later.
In ref.~\cite{ref:Davanco_BE}, the characteristics of the X$_1$ and
X$_2$ emission lines were investigated through pump-power-dependent
and temperature-dependent PL measurements.  Here, we focus on the
single photon nature of the emission lines by measuring photon
statistics through the HBT interferometer setup
(Fig.~\ref{fig:Figure3}(b)) under similar excitation conditions as in
the previous work. Figure~\ref{fig:Figure5}(b) shows the result
of an intensity autocorrelation measurement performed on the
spectrally filtered $X_{1}$ line at its saturation power. A clear
suppression of the peak at zero time delay to a value of
$g^{(2)}(\tau)\,=\,0.37\,<\,0.5$ is
seen~\cite{ref:bullseye_followup_note_1}, verifying that the
measured line is originated from a single QD \cite{ref:Michler}. The
deviation from the ideal value of $g^{(2)}(\tau)\,=\,0$ for a single
quantum emitter is most likely related to the uncorrelated
background emission coupled to the cavity mode at the same
frequency.

\begin{figure}[t]
\centerline{\includegraphics[width=0.95\linewidth]{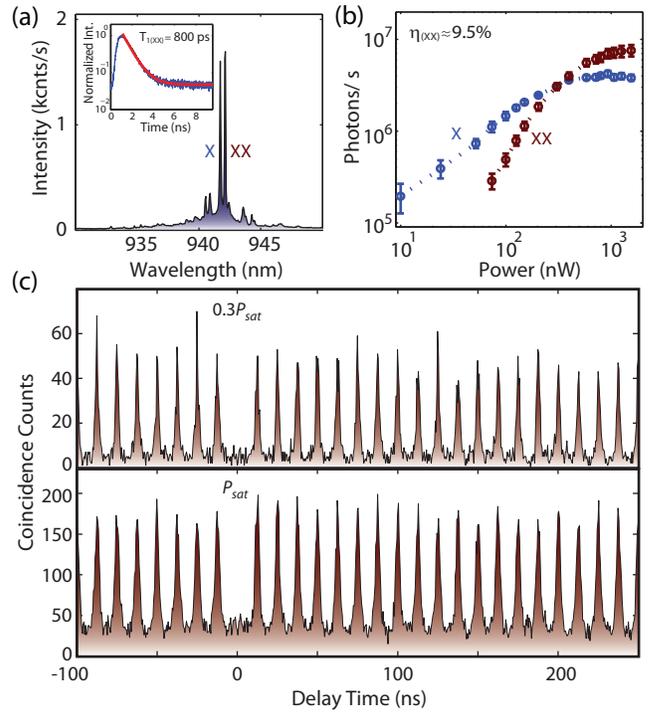}}
 \caption{Bright single-photon generation from a single QD coupled to the FM of the cavity.
a) Low temperature (T\,=\,10\,K) $\mu$-PL spectrum of device BE2.
Two bright emission lines are visible and labeled as exciton (X) and
bi-exciton (XX) based on their excitation power dependence as shown
in b). inset: Time-resolved PL measurement on XX emission line,
which reveals $T_1\,=\,800$~ps. c) Demonstration of single-photon generation from the XX emission
under low power $P\,=\,0.3P_{sat}$ (upper panel) and high power
$P\,=\,P_{sat}$ (lower panel) conditions.}
\label{fig:Figure6}
\end{figure}

In Fig.~\ref{fig:Figure5}, bright single photon generation from a
single QD with a large collection efficiency and fast fluorescence
decay time have been demonstrated. However, the autocorrelation
measurements resulted in a clear background at the QD emission
frequency due to contributions from other states resonant with the
cavity mode emission. This suggests a potential trade-off between
either (or both) high extraction efficiency and single photon purity
of the collected signal or fast fluorescence decay time and purity
of the collected signal. To further investigate this, two more
devices (named as BE2 and BE3) were studied in detail.
Figure~\ref{fig:Figure6}(a) shows a low power emission spectrum of
device BE2 taken under above-band pulsed excitation of the GaAs at
T\,=\,10\,K. Similar to the device BE1, this device also shows two
bright states named as excitonic (X) and bi-excitonic (XX) emission,
which sit on top of the FM of the cavity around $942$~nm.
Fig.~\ref{fig:Figure6}(b) depicts the output intensity of X and XX
emission lines as a function of excitation power in a
double-logarithmic plot~\cite{ref:bullseye_followup_note_2}. As is
seen clearly, the X line has a linear power dependence while the XX
line shows a quadratic increase with power as expected for exciton
and bi-exciton emission from a single QD,
respectively~\cite{ref:Thompson2002}. The output intensity of the X
and XX emission lines in the figure are given as number of photons
collected at the microscope objective, thus providing a direct
estimate of total collection efficiency of the corresponding
emission. Conversion of the integrated intensity of CCD counts to
number of photons was done by measuring the total transmission of
the optical path, as discussed in detail in the supplementary part
of \cite{ref:Davanco_BE}. The extraction efficiency of exciton and
bi-exciton emission are estimated as 5.4~$\%\pm~0.4~\%$ and
9.5$~\%\pm~1.4~\%$, respectively, by comparing the number of photons
collected at the saturation power to the total number of photons
generated from the corresponding emission line, assuming 100~$\%$
radiative efficiency of the QD.

\begin{figure}[t]
\centerline{\includegraphics[width=0.95\linewidth]{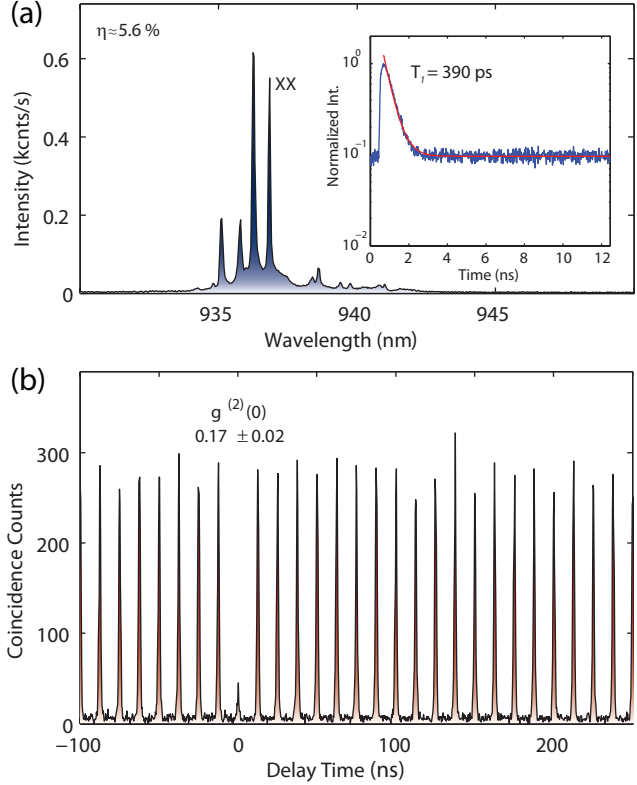}}
 \caption{Purcell-enhanced bright single-photon generation from a single QD.
(a) Low temperature (T\,=\,10\,K) $\mu$-PL spectrum of device BE3,
which shows sharp emission lines around 936\,nm within the spectral
width of the fundamental mode of the bullseye. inset: Time-resolved
PL on QD emission at 937\,nm. (b) The result of an autocorrelation
measurement performed on XX line under close-to-saturation
excitation power.} \label{fig:Figure7}
\end{figure}

Having shown photon collection efficiencies on par with those seen
in device BE1, we also performed autocorrelation measurements on the
XX line under weak ($P\,=\,0.3P_{Sat}$) and strong excitation powers
($P\,=\,P_{Sat}$), the results of which are shown in
Fig.~\ref{fig:Figure6}(c). Almost complete suppression of the
$\tau\,=\,0$ peak is observed even at the saturation power, yielding
a great improvement compared to the results obtained from device
BE1. The constant background observed in the autocorrelation
measurements is mainly due to the overlap of broad correlation peaks
indicating long lifetime of the measured emission. The inset of
figure~\ref{fig:Figure6} (a) shows a direct measure of the lifetime
of the XX emission through time-resolved PL setup. Applying an
exponential fit to the measured decay curve gives the radiative
lifetime as $T_1\,=\,800\,\pm\,70$~ps, slightly more than a factor of two
larger than that measured for device BE1.

The results obtained from BE2 showed a clear improvement in the
purity of the single photons generated from a single QD, with a cost
of a longer radiative lifetime. A compromise between these two
important parameters can also be achieved.
Figure~\ref{fig:Figure7}(a) shows a low temperature $\mu$-PL
spectrum of another device, named BE3, under weak pulsed excitation
($\lambda_{Exc.}=820$~nm). Like the other devices, bright emission
lines are visible on top of the cavity mode around $936$~nm, and are
collected with an efficiency of $5.6~\%\pm0.4~\%$ at saturation (not
shown). The inset of Fig.\ref{fig:Figure7}(a) depicts the result of
a time-revolved PL measurement performed on the XX line, which
dominates the spectrum at elevated powers, and a radiative lifetime
of about 390$~\pm$15~ps is estimated from the bi-exponential fit.
This radiative lifetime is close to that measured for device BE1 and
about a factor of two faster than that for BE2. In order to verify
the single-photon nature of the measured emission line, an
autocorrelation measurement was performed, the result of which is
shown in Fig.~\ref{fig:Figure7}(b). The reduced area of the peak at
zero time delay indicates a clear photon antibunching with a value
of $g^{(2)}(\tau)\,=\,0.17\,<\,0.5$, thus proving that the source is
dominantly composed of single photons with a multi-photon
probability in-between that of devices BE1 and BE2.

Devices BE1, BE2, and BE3 show radiative lifetimes of 360 ps, 800
ps, and 390 ps, indicating that the relative level of Purcell
enhancement varies by a factor of 2.2 between the different
devices. An estimate of the absolute Purcell enhancement factor
$F_{p}$ requires a measurement of the QD lifetime without
modification by the cavity. Unfortunately, the broad spectral
bandwidth of the bullseye eliminates the potential for using
temperature tuning to shift the QD lines off-resonance from the
cavity.  We instead measured the lifetime of QDs within suspended
waveguides~\cite{ref:Davanco_WG} made from the same wafer, where we
saw characteristic lifetimes between 1.4 ns and 1.6 ns for the
neutral exciton (X) state (no Purcell enhancement is expected in
these devices).  However, as described above, the lifetime and
$g^{(2)}(\tau)$ measurements for the bullseye devices were typically
performed on the biexciton (XX) state, though we note that the
collection efficiency of the exciton state was often equally as
high. While the XX lifetime is in principle a factor of two times
shorter than the X lifetime, other works~\cite{ref:Santori5,ref:Kiraz3}
have suggested that quantum confinement properties specific to the particular QD geometry may
limit this reduction to a factor closer to 1.4. Taking this into
account, we roughly estimate that a Purcell factor as high as
$\approx$3 has been demonstrated in these devices.

\subsection{Coherence time}

For quantum information processing applications, it is important
not only to suppress multi-photon events but also to have the single photon
wavepackets be indistinguishable~\cite{ref:Kiraz,ref:Santori2}.  For
an ideal transform-limited QD line, the coherence time $T_{2}$ would
be twice the lifetime $T_{1}$, and values approaching this limit have been
demonstrated~\cite{ref:Bayer,ref:Ates_PRL09}. In practice,
achieving such coherence times depends strongly on parameters such
as temperature and the method by which the QD is optically excited
(e.g., above-band vs. resonant excitation).

We investigate the coherence properties of the bright emission lines
from all three bullseye devices by measuring the first-order field
correlation function $g^{(1)}(\tau)$, which provides a direct access
to the coherence time of the emission through \cite{ref:Loudon2000}:
\vspace{4 mm}
\begin{equation}\label{eq:tau}
\tau_c = \int_{-\infty}^{\infty}|g^{(1)}(\tau)|^2 d\tau
\vspace{4 mm}
\end{equation}

\begin{figure}[t]
\centerline{\includegraphics[width=0.95\linewidth]{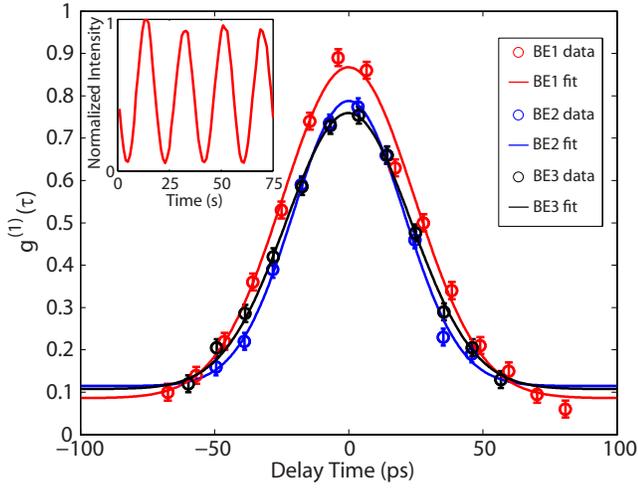}}
 \caption{The results of first-order field correlation function measurements on
all three devices under low power above-band excitation conditions.
The inset shows an example of the interference fringes produced at
a position close to zero delay line in the Michelson interferometer. }
\label{fig:Figure8}
\end{figure}

Experimentally, the $g^{(1)}(\tau)$ function is measured from the
visibility of the interference fringes observed at the output port
of the Michelson interferometer (Fig. \ref{fig:Figure3}(d)) according
to: \vspace{4 mm}
\begin{equation}\label{eqn:visiblity}
V(\tau)~=~\frac{(I_{max}~-~I_{min})}{(I_{max}~+~I_{min})}~=~\left|
g^{(1)}(\tau) \right|
\vspace{4 mm}
\end{equation}
\\
for equal intensities of interfered light. The inset to
figure~\ref{fig:Figure8} shows an example of high-resolution interference
fringes of the XX emission line from device BE1 obtained under
above-band excitation conditions ($\lambda_{Exc.}=780$~nm, CW). The
visibility of the fringes at each systematically varied time delay
between the interferometer arms are calculated according to
Eqn.~\ref{eqn:visiblity}, and plotted in Fig.~\ref{fig:Figure8} for all
devices~\cite{ref:bullseye_followup_note_3}. The visibility of the
interference fringes decreases with an increase in delay as a result
of the limited coherence time $\tau_c$ of the emission. Applying a
Gaussian fit to the visibility data reveals $\tau_c\,=\,27$\,ps
($\Gamma\,=\,100\,\mu$eV) for BE1 and 17\,ps (160\,$\mu$eV) for BE2
and BE3. The emission linewidth estimated from $g^{(1)}(\tau)$
measurements is in full agreement with the linewidth obtained from
the corresponding $\mu$-PL spectrum for BE2 and BE3. However, the
emission spectrum of device BE1 revealed a linewidth of
130\,$\mu$eV, which is broader than the linewidth obtained from
$g^{(1)}(\tau)$ measurement (100\,$\mu$eV). The difference mainly
arises from the limited spectral resolution of the $\mu$-PL setup
estimated as 80\,$\mu$eV by measuring a narrow CW laser. The
measured linewidths of emission from all devices are much broader
than the reported homogenous linewidth of QDs that are nearly
transform-limited ($\approx1\mu$eV to 2
$\mu$eV)~\cite{ref:Bayer,ref:Ates_PRL09}. The main reason for this
is thought to be the incoherent nature of the above-band excitation
process, which excites several QDs within the excitation spot and
thus enhances the dephasing processes due to the carrier-carrier
interactions. In addition to that, spectral diffusion of the
emission frequency due to fluctuating charges in the vicinity of the
QDs may result in a broadening of the emission
\cite{ref:Seufert2000}.

\section{Discussion}
\label{sec:discussion} The results presented above can be further
interpreted with the help of numerical simulations.  In particular,
we try to understand how the specific device geometry influences the
aforementioned trade-offs in different performance characteristics,
as well as the reduced collection efficiency and rate enhancement
observed in experiments relative to the maximum predicted
theoretical values in ref.~\cite{ref:Davanco_BE}.

\subsection{Tradeoffs in Purcell enhancement and collection
efficiency} \label{subsec:sim_discuss}

\begin{figure}[t]
\centerline{\includegraphics[width=8.5cm]{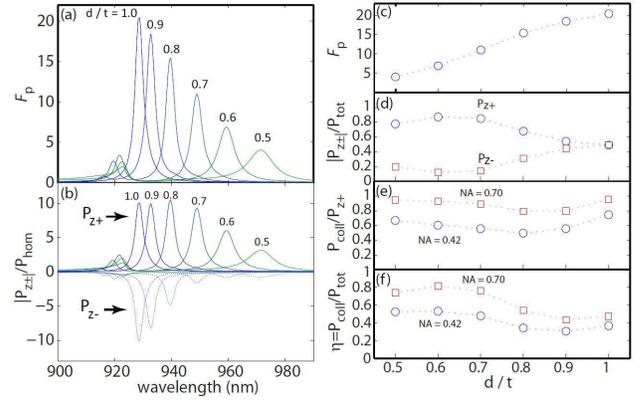}}
\caption{(a) Purcell enhancement factor $F_p$ as a function of
wavelength for various normalized trench depths $d/t$. (b) Ratio of
the vertically emitted power in the $\pm z$ direction ($P_{z\pm}$)
to the total emitted power of a dipole in bulk GaAs
($P_{\text{hom}}$), as a function of wavelength for various values
of $d/t$. (c) $F_p$ as a function of $d/t$. (d) percentage of total
power emitted upwards and downwards, as a function of $d/t$. (e)
percentage of upwards emitted power collected by NA = 0.42 NA = 0.7
lenses. (f) percentage of total emitted power collected by NA = 0.42
NA = 0.7 lenses. \label{fig:Figure9}}
\end{figure}

We next illustrate the effects of varying trench depths on emission
properties of the circular dielectric grating. The first key
quantity of interest is the Purcell enhancement factor $F_p$. It is
determined from simulation by the quantities $P_{\text{tot}}$ and
$P_{\text{hom}}$.  $P_{\text{tot}}$ is the total power emitted by a
dipole in the bullseye structure, while $P_{\text{hom}}$ is the
power emitted by a dipole in a homogenous GaAs medium, so that $F_p$
is given by their ratio
$P_{\text{tot}}/P_{\text{hom}}$~\cite{ref:Xu_99}.
Figures~\ref{fig:Figure9}(a) and (c) show $F_p$ as a function of
wavelength and normalized trench depth $d/t$. $F_p$ increases as
$d/t$ approaches unity, with deeper trenches providing increased
field confinement and higher cavity $Q$. This is a consequence of
better overlap of the guided field inside the slab and the etched
region, which leads to increased guided wave reflectivity and
reduces coupling to out-of-plane radiation.  The higher cavity $Q$
with increasing trench depth is seen in the decreasing resonance
width in Fig.~\ref{fig:Figure9}(a); in addition, a strong
blueshift of the cavity mode is observed.

The increase in $F_p$ with trench depth comes at a cost, however, as
the fraction of light emitted above the structure ($+z$ direction)
is reduced.  Intuitively, this makes sense in that, once the grating
is completely symmetric ($d/t$=1, for which $F_p$ is maximized),
equal emission in the $+z$ and $-z$ directions should be expected.
Figures~\ref{fig:Figure9}(b) and (d) show the upwards
($P_\text{z+}$) and downwards ($P_\text{z-}$) extracted powers as a
function of wavelength and normalized trench depth $d/t$. Asymmetric
and preferential upwards emission (determined by the ratio
$P_{z+}$/$P_{z-}$) is maximized with $d/t=0.6$, for which
$P_{z+}/P_{z-}=7.1$, but comes at the expense of a significantly
reduced $F_p$ of 6.9 in comparison to its peak value of nearly 20.

For useful single photon emission, preferential upwards emission
alone is not sufficient. We also require the emission to be directed
within a relatively narrow divergence angle far-field pattern, so
that the majority of this emission can be collected by a standard
focusing optic.  Figure~\ref{fig:Figure9}(e) shows the fraction of
upwards emission that is collected
($P_{\text{coll}}$/$P_{\text{z+}}$) by NA=0.42 and NA=0.7 optics, at
the peak emission wavelength for each $d/t$ (see
Fig.~\ref{fig:Figure9}(a)). We note that this value is close to
unity for the NA=0.7 optic, indicating a strong level of
directionality, which we also expect based on the far-field emission
pattern in Fig.~\ref{fig:Figure2}(d). In addition, we see that the
collection of the upwards-emitted power can vary by as much as
33~$\%$ for the trench depth range considered. Finally,
Fig.~\ref{fig:Figure9}(f) shows the overall collection efficiency
$\eta$ (=$P_{\text{coll}}$/$P_{\text{tot}}$) as a function of trench
depth, which is the quantity that is actually measured in the
experiments of the previous section. $\eta$ peaks at a value of
$\approx$80~$\%$ for a NA=0.7 optic, while for the NA=0.42 optic
used in our experiments, this value is $\approx$50~$\%$.

It is evident from these results that sufficiently accurate control
of the trench depth is a necessity, not only for optimal spectral
alignment, but also for optimal extraction efficiency. For instance,
for a trench depth $d=0.8t$, the collection efficiency into a 0.42
NA objective drops to approximately 34~\% from the 53~\% for
$d=0.6t$. As $t=190$~nm, a 40~nm difference in trench depth, which
we might expect to observe based on the tolerances of our
fabrication process, can lower the maximum extraction efficiency
significantly.

\begin{figure}[t]
\centerline{\includegraphics[width=8.5cm]{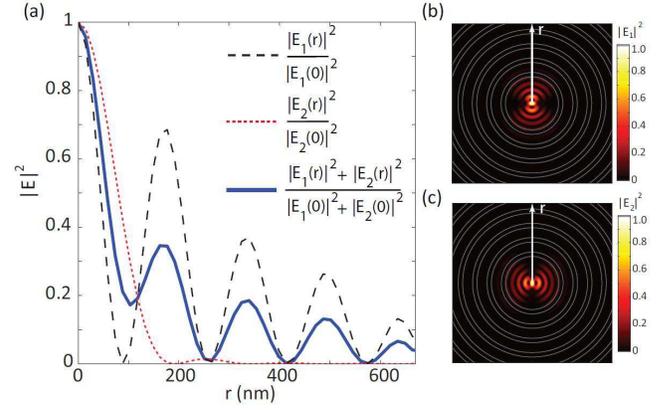}}
\caption{(a) Electric field squared of first order modes
$\mathbf{E_{1}(r)}$ and $\mathbf{E_{2}(r)}$ , with $r$ the distance
along the radial direction. Continuous line: sum of squared electric
fields for modes $\mathbf{E_{1}}$ and $\mathbf{E_{2}}$. Field
profiles for modes $\mathbf{E_{1}(r)}$ and $\mathbf{E_{2}(r)}$,
which are excited by (b) $x$-polarized and (c) $y$-polarized
dipoles, respectively.}\label{fig:Figure10}
\end{figure}

Along with radiative lifetime (determined by $F_p$) and collection
efficiency, the third key characteristic of the devices measured in
the previous section is the single photon purity (i.e., the value of
$g^{(2)}(0)$).  Our simulations do not directly address the effects
which cause $g^{(2)}(0)>0$. However, many
works~\cite{ref:Winger2009,ref:Laucht_Finely_2010} have attributed
this to cavity-enhanced out-coupling of multi-excitonic and
hybridized QD-wetting layer states that are spectrally resonant with
the QD state of interest.  In such a scenario, strongly
Purcell-enhanced devices are likely to have $g^{(2)}(0)$ levels that
significantly differ from zero, while devices lacking Purcell
enhancement can show $g^{(2)}(0)\approx0$ even under above-band
pumping near saturation~\cite{ref:Claudon}. From this, we speculate
that shallower grating depths ($0.5<d/t<0.6$), which lead to smaller
$F_p$ in Fig.~\ref{fig:Figure10}(c), may also lead to enhanced
emission purity. Furthermore, this benefit would come together with
increased collection efficiency  (Fig.~\ref{fig:Figure10}(f)) as a
consequence of improved vertical extraction and emission asymmetry
(Fig.~\ref{fig:Figure10}(d)).

\subsection{Influence of Dipole Position and Orientation}

Other important factors affecting both the achieved Purcell
radiative rate enhancement and overall photon collection efficiency
are dipole position and orientation within the nanostructure.
Considering the fundamental mode of the BE cavity in
Fig.~\ref{fig:Figure2} (shown also in Fig.~\ref{fig:Figure10}(b)
and (c)), which we label as $M_0$, we note that due to the circular
symmetry of the nanostructure, both mode $M_0$ and a degenerate,
90$^{\circ}$ version of it (henceforth referred to as $\mathbf{E_1}$
and $\mathbf{E_2}$, respectively) are simultaneously supported. Mode
$\mathbf{E_1}$ is exclusively excited by azimuthally oriented
dipoles along the radial direction ($\mathbf{r}$), while mode
$\mathbf{E_2}$ is exclusively excited by radially oriented dipoles
along $\mathbf{r}$. The specific instances of the modes excited by
$x$-polarized and $y$-polarized dipoles are shown in
Fig.~\ref{fig:Figure10}(b) and (c), and a dipole oriented at an
arbitrary angle in the plane would then be expected to radiate into
both of these modes.

Figure~\ref{fig:Figure10}(a) shows the intensities of
$|\mathbf{E_{1}}|^2$ and $|\mathbf{E_{2}}|^2$ as functions of
position $r$ along the radial direction (in this case, the $y$
direction), relative to those at the center. Since the radiative
rate into a particular mode is proportional to the local modal
electric field squared~\cite{ref:Xu_99}, it is apparent that Purcell
radiative rate enhancement can vary significantly over very small
position ranges, and also with dipole orientation. For a dipole with
equal radial and azimuthal components, the plot of
$|\mathbf{E}({r})|^2=|\mathbf{E_{1}}(\mathbf{r})|^2+|\mathbf{E_{2}}({r})|^2$
shown in Fig~\ref{fig:Figure10}(a) makes it apparent that
positioning within a $\approx200$~nm radius from the center is
necessary for maximal modal coupling and Purcell enhancement to be
achieved.

\begin{figure}[t]
\centerline{\includegraphics[width=8.5cm]{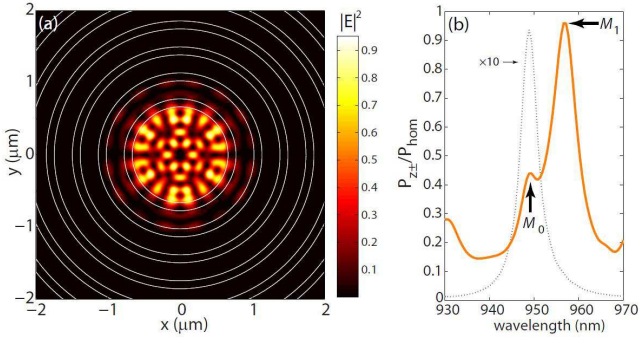}}
\caption{(a) Field profile of a higher order mode of the structure
in Fig.~\ref{fig:Figure2}(a), excited by a horizontally oriented
dipole located 260~nm away from the center. (b) Upwards-emitted
power, normalized to the homogeneous medium dipole radiated power,
for the offset dipole. Dotted line: same, for dipole at the center;
note the 10x scaling between the two curves.} \label{fig:Figure11}
\end{figure}

Coupling to both of the degenerate $M_{0}$ modes is not the only
potential source of non-ideality for dipoles located away from the
center of the cavity. Due to its large dimensions, the cavity
supports not only the mode shown in Fig.~\ref{fig:Figure2}(a) and
(b), but also an ensemble of broad, spectrally overlapping
resonances that can be excited by dipoles offset from the bullseye
center. These resonances can have $\beta$-factors comparable to that
of the main mode (depending on the dipole location), however less
directional far-fields. In this situation, an enhanced spontaneous
emission rate can be achieved (with contributions from all modes),
together with a reduced overall collection efficiency. To
illustrate, we show in Fig.~\ref{fig:Figure11}(a) the field profile
of a high order mode that is excited on the same grating as in
Fig.~\ref{fig:Figure2}(a), when an $x$-oriented dipole is placed
at a distance of 260 nm away from the center. As shown in
Fig.~\ref{fig:Figure11}(b), this broad mode ($M_1$) is centered at
957~nm, and overlaps with the main cavity mode $M_0$ at 949~nm
(compare with $P_{z+}$ spectrum obtained for a centered dipole shown
by the dotted line). The far-field of mode $M_1$ is such that only
$\approx26$~\% of the total upwards radiated power can be collected
by an NA=0.42 objective, significantly smaller than the
$\approx56$~\% for $M_0$. It is also worthwhile noticing in
Fig.~\ref{fig:Figure11}(b) that the radiative rates for both modes
M$_0$ and M$_1$ for the offset dipole case are considerably lower
than for a centered dipole (dotted line). This is expected because
the modified radiative rate for a dipole in the nanostructure is
proportional to $|\mathbf{E}(\mathbf{r_0})|^2$, where
$\mathbf{E}(\mathbf{r_0})$ is the electric field at the dipole
position and for both modes the offset dipole is located near a
null.

\subsection{Quasi-Resonant Excitation}
\label{subsec:quasi_resonant} 
\begin{figure}[t]
\centerline{\includegraphics[width=0.95\linewidth]{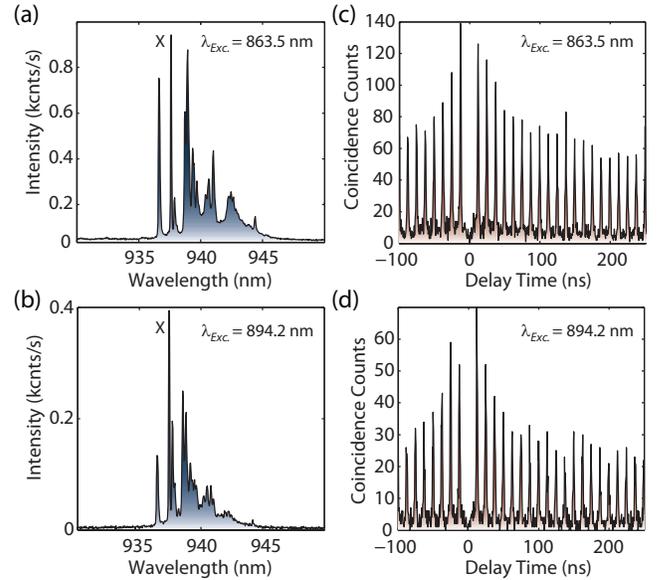}}
 \caption{Higher-energy state excitation of a single QD in device BE1. (a)
Emission spectrum of device BE1 under quasi-resonant excitation with
$\lambda_{Exc.}\,=\,863.5$~nm and (c) measured second-order
correlation function of excitonic emission line X at $937$~nm. The
vanished peak at zero time delay indicates a strong suppression of
multi-photon emission probability. (b) Emission spectrum of BE1 with
$\lambda_{Exc.}=894.2$~nm and (d) the measured second-order
correlation function of the same peak.} \label{fig:Figure12}
\end{figure}

The coherence properties of single QD
emission can be improved under selective excitation conditions,
where the laser energy is tuned to a higher energy state of an
individual QD. This so called "quasi-resonant" excitation process
limits the number of excited QDs, thus reducing the dephasing
processes causing a linewidth broadening as mentioned before.
Moreover, it also reduces the multi-photon generation probability
significantly because of a reduced probability of creating more than
one electron-hole pair in the QD. Figures\,\ref{fig:Figure12}(a)
and (b) show emission spectra of device BE1 taken under
quasi-resonant excitation, where the pulsed laser is tuned to
863.5~nm and 894.2~nm, respectively. In each spectrum, a bright
excitonic emission line is visible at 937~nm, on which an
autocorrelation measurement was also performed to investigate its
single emitter nature. The result of the experiments are depicted in
Figures\,\ref{fig:Figure12} (c) and (d). A complete missing peak
observed at zero time delay in both figures proves that the applied
excitation scheme resulted in almost perfect single-photon
generation from the measured emission line. An interesting feature
seen in the figures is that the correlation peaks close to zero
delay are much larger than the outer side peaks. Similar results
were reported under quasi-resonant excitation of a single QD
\cite{ref:Santori4} and the observed blinking effect on a long time
scale was attributed to a charge fluctuation of the QD. The
linewidth of the measured state is extracted as
$\Gamma\,=\,100\,\mu$eV from the corresponding $\mu$-PL spectrum,
which is narrower than the typical linewidths measured in $\mu$-PL
measurements under above-band excitation conditions.

\section{Conclusions}
\label{sec:conclusions}

We have presented a series of measurements of single quantum dots in
circular Bragg grating 'bullseye' microcavities.  These devices,
first presented in ref.~\cite{ref:Davanco_BE}, offer a combination
of features valuable to single photon sources: high extraction
efficiency, moderate spectral bandwidth, and radiative rate
enhancement, all within a planar geometry requiring a relatively
straightforward fabrication procedure.  Here, we focus on the
tradeoff between rate enhancement and multi-photon probability
present in these devices under above-band optical pumping. We first
demonstrate that the brightest devices studied in
ref.~\cite{ref:Davanco_BE} indeed show antibunched photon
statistics, but that the multi-photon probability is somewhat high
(though still low enough that the source is dominantly comprised of
single photons).  We then consider two other devices in which the
collection efficiency remains high, but the multi-photon probability
is reduced, albeit at the expense of a longer radiative rate.  We
find that a near-complete suppression of multi-photon probability is
possible in devices with high collection efficiency, but that the
rate enhancement is correspondingly the lowest we see. These results
suggest that unwanted background radiation (for instance, from
transitions of multi-excitonic or hybrid quantum well-QD states) is
enhanced and funneled through the cavity, leading to less pure
single photon emission~\cite{ref:Winger2009,ref:Laucht_Finely_2010}.
Cavities with reduced Purcell enhancement may thus lead to reduced
multi-photon emission~\cite{ref:Claudon}.

Numerical simulations were performed to help interpret the physical
scenarios that lead to the aforementioned compromises and the
reduced collection efficiency and rate enhancement values observed
relative to the maximum predicted values. The most straightforward
method to improving the observed collection efficiency is through
using an increased numerical aperture collection optic.  Moving to
an NA\,=\,0.7 optic from the current NA\,=\,0.42 optic is predicted
to improve the collection efficiency for existing devices by
$\approx$\,60~$\%$.  Beyond this, precise control of the device
geometry, both in terms of the etched trench depth and the QD
position~\cite{ref:Hennessy3}, are needed to achieve the desired
combination of rate enhancement and collection efficiency. In
particular, shallower grating trenches lead to increased light
outcoupling, as well as a more highly asymmetric vertical emission,
which ultimately leads to higher collection efficiencies. This comes
at the cost of lower Purcell enhancement, which however may
potentially lead to a more pure single photon emission, as discussed
above.

Ideally, the bullseye cavity would be useful not only as a bright
single photon source but as a bright source of indistinguishable
single photons.  Measurements under above-bandgap excitation
indicate that the QD coherence time is extremely short ($<$\,30 ps),
so that quasi-resonant excitation~\cite{ref:Santori2,ref:Ates_PRL09}
or post emission spectral/temporal
filtering~\cite{ref:Patel_PRL08,ref:Rakher_APL_2011} will be
required to generate indistinguishable photons.  As a preliminary
step towards this, we present photoluminescence spectra and photon
correlation measurements of devices optically pumped at a wavelength
resonant with higher states of the QD and/or cavity, for which
near-complete suppression of multi-photon events is observed.  Such
quasi-resonant excitation may be an eventual route to simultaneously
achieving high collection efficiency, strong rate enhancement, and
suppressed multi-photon probability.

\section{Acknowledgements}

This work has been partly supported by the NIST-CNST/UMD-NanoCenter
Cooperative Agreement. We thank Nick Bertone from OptoElectronic
Components for loan of a high timing resolution single photon
detector.  The identification of any commercial product or trade
name does not imply endorsement or recommendation by the National
Institute of Standards and Technology.

\bibliography{./Ates_Bullseye}

\begin{thebibliography}{10}
\bibitem{ref:bullseye_followup_note_1}
The $g^{(2)}(0)$ value is determined by comparing the integrated area of the
  peak around time zero to the average area of the peaks away from time zero.
  The uncertainty on this value is given by the standard deviation in the area
  of the peaks away from time zero.
\bibitem{ref:bullseye_followup_note_2}
The photon count rate vs. pump power plot in Fig. 6(b) is determined by fitting
  the spectral peak to a Lorentzian and integrating the area within the
  Lorentzian. Error bars are due to the uncertainty of the fit and are one
  standard deviation in amplitude.
\bibitem{ref:bullseye_followup_note_3}
The error bars in the $g^{(1)}(\tau)$ data in Fig. 8 are determined by the
  fluctuations in the inteference contrast and are one standard deviation in
  amplitude.
\end{thebibliography}
\bibliographystyle{IEEEtran}

\end{document}